\documentclass[12pt,a4paper,final]{iopart}

\usepackage{iopams}  
\usepackage{graphicx}
\usepackage{feyn}

\begin{document}

\title[Neural Networks in Hadron Physics]{Applications of Neural Networks in Hadron Physics}

\author{Krzysztof M. Graczyk and Cezary Juszczak}
\address{Institute of Theoretical Physics, University of Wroc\l aw,\\
pl. M. Borna 9,
50-204, Wroc\l aw, Poland}
\ead{krzysztof.graczyk@ift.uni.wroc.pl}

%\author{Cezary Juszczak}
%\address{Address Three, Neverland}
%\ead{author.two@mail.com}

\begin{abstract}
The Bayesian approach for the feed-forward neural networks is reviewed. Its potential for usage in hadron physics is discussed. As an example of the application the study of the the two-photon exchange effect is presented. We focus on the model comparison, the estimation of the systematic uncertainties due to the choice of the model, and the over-fitting. As an illustration the predictions of the cross sections ratio $d \sigma(e^+ p\to e^+ p)/d \sigma(e^- p\to e^- p)$ are given together with the estimate of the  uncertainty due to the parametrization choice.
\end{abstract}

\vspace{2pc}
\submitto{\JPG}
%\maketitle

\section{Introduction}

One of the goals of physics is to  construct models, which describe a part of reality. A promising model should be able to reproduce the experimental data with \textit{reasonable precision} and should also be characterized by good predictive power.  The closest to Nature seems to be a theory, which  is based on fundamental symmetries or some other beautiful mathematical structure  and contains a minimal number of  internal parameters. However, in many cases either  the fundamental underlying theory is not known yet or the model is not fully solvable  yet. Therefore effective approaches are often utilized to describe physical observables and properties. They are defined by a set of internal parameters usually inferred from the measurements. Some of them, like particle masses, have particular physical interpretation  but many are just introduced  to reproduce the experimental data. 

Modelling the internal structure of the nucleon is an example of a situation where  the theory, 
at least in some regions, is unsolvable or very difficult to apply. 
Because of the asymptotic freedom, the perturbative methods in quantum chromodynamics (QCD)  work well at large energies, but they fail in the confinement region. 
In this low energy range it is more convenient to describe the system   in terms of hadronic degrees of freedom (baryons and mesons) rather than the quarks and gluons.
Therefore the internal hadronic structure, in the confinement region, is usually investigated within effective approaches. 
In many of them the information about the static and dynamical  internal structure \cite{Thomas_book} of hadrons is parametrized by the transition form factors (FFs). 
In the case of the nucleon they describe its electromagnetic (E-M) as well as electroweak properties.
Their functional form is not known and  they are inferred from the scattering data\footnote{We notice that many efforts have been made to calculate the FFs within the lattice QCD \cite{latice}.}.

Usually in the particle and nuclear physics the methods of frequentistic rather than Bayesian statistics are used.  There are fundamental differences between both methodologies starting from  the very  definition of probability (for comprehensive review see \cite{Jeffreys,De_Agositini}). In the Bayesian approach the probability is \textit{the measure of the degree of belief that an event will occur} \cite{De_Agositini}. Seemingly this definition is subjective and non-operational in contrast to the frequentistic approach, where the probability is defined by \textit{the ratio of the number of times the event occurs in a test series to the total number of trials in the series}. The latter definition implies an additional assumption that every event occurred/occurs/will occur with the same probability \cite{De_Agositini}. In the Bayesian statistics, with the use of the Bayes' rule, one can construct the probability (posterior), which accommodates the initial model assumptions  (prior and likelihood) with the data. The posterior should always be updated after new data arrive. The  statistical model (see definition in the next section) is defined by the probability distribution of its parameters and the model assumptions contained in the definition of the prior and the likelihood. 

Having a set of physical hypotheses (models) it is natural to ask:  \textit{which one is the most favourable by given data?}  Within the Bayesian statistics the hypothesis can be ranked by the conditional  probability $P({\rm hypothesis}|{\rm data})$. 
Therefore the comparison of different models and the discussion of the 
 impact of the initial assumptions on the results of the analysis can be naturally performed.
 Moreover the analysis of every possible model brings a valuable contribution. 
 Indeed even negative verification of a particular hypothesis is constructive information which contributes to the posterior needed to classify the hypotheses.  
 
It is  believed that the laws of Nature are simple, therefore the desired theory, which aims to approach the \textit{true underlying theory}, should be based on a small set of fundamental assumptions and it should contain a minimal number of  internal parameters. Hence it is rather natural to search for simpler rather than more complex descriptions of the physical reality. An instructive example is the extraction of the value of the proton radius from the elastic $ep$ scattering data.  This quantity  is  related with the slope of the electric proton form factor ($G_{E}$) at vanishing four-momentum  transfer $Q^2\to 0$. In the typical analysis  the parametrizations for the electric $G_{E}$ and magnetic $G_{M}$ proton form-factors are postulated. Usually these are arbitrary functions, which obey some general properties and they  are fitted to the experimental data. It turns out that  the obtained value of the proton radius depends on the choice of the parametrization \cite{Hill_et_al}. This difficulty can be approached within the Bayesian statistics \cite{Graczyk:2014lba}, which  in natural way embodies the Occam's razor principle \cite{Occam_razor} (models with lower number of parameters are preferred).

In this paper we shall introduce a statistical framework, based on the Bayesian statistics, which allows to quantitatively control the model-dependence of predictions of physical quantities, and   to estimate the systematic uncertainties caused by a particular choice of the model.  
 The proper estimate of the statistical and systematic uncertainties is of importance in atomic physics \cite{theeditors_of_PRA}, nuclear physics \cite{Dobaczewski:2014jga} but also in the physics of hadrons.

There are a lot of lepton-hadron and hadron-hadron scattering cross section measurements. The analysis of these data brings information about the internal structure of hadrons and allows for  validation of the theoretical models. With the help of the neural network methods one can try to analyse these data in a model-independent way, constructing the statistical model, based on which  the predictions about the transition FFs \cite{Graczyk:2010gw,Graczyk:2011kh} and the parton distribution functions \cite{Ball:2013lla} can be made. 

In the following sections we shall introduce the Bayesian framework (BF) for feed-forward neural networks and as an instructive example of application we will present results of our  studies of the proton structure, E-M FFs and the two-photon exchange (TPE) effect \cite{Graczyk:2013pca,Graczyk:2011kh}.
Investigation of the proton FFs and related observables (proton radius, two photon exchange effect) is an important topic of the hadron physics \cite{Thomas:2014zja}.  In this paper we  discuss the statistical features of the framework concentrating our attention on the quantitative model comparison and the estimate of the systematic uncertainties due to the choice of initial model assumptions.

The paper is organized as follows. Sect. 2 introduces the BF for feed-forward neural networks. In Sect. 3 the application to hadron physics  is presented. In Sect. 4 the features of the approach and the results are discussed.

\section{Remarks on the Bayesian Framework}

In this section we recall, following D.\ MacKay \cite{MacKay_thesis}, some general features of the BF.

Our purpose is to find the optimal  model or a set of  models having a given set of measurements. By a statistical model we mean: 
\begin{itemize}
\item[(i)] the  function, $\mathcal{N}$,  used to fit the data;
\item[(ii)] two conditional probabilities: the distribution of the function parameters $P(\{w_i\}|\mathcal{N})$ and the likelihood $P(\mathcal{D}|\{w_i\}, \mathcal{N})$,
where
$\mathcal{D}$ denotes the data, while $\{w_i\}$ are the model parameters.
\end{itemize} 

In principle one should consider all possible hypotheses, for each of them find the most optimal set of parameters    and  rank them by the conditional  probability $P(\mathcal{N}|D)$, which estimates how plausible any given hypothesis is according to the measurements.

The optimal configuration of the parameters, $\{w_i \}_{MP}$, of the model maximizes the posterior  (obtained from the Bayes' rule):
\begin{equation}
\label{posterior_paraeters_general}
P(\{w_i\}| \mathcal{D},\mathcal{N}) = \frac{P(\mathcal{D}|\{w_i \}, \mathcal{N})P(\{w_i \}|\mathcal{N})}{P(\mathcal{D}| \mathcal{N})},
\end{equation}  
where $P(\mathcal{D}|\{w_i \}, \mathcal{N}) $ is the likelihood and $P(\{w_i \}|\mathcal{N})$ is the prior which contains information about the initial assumptions. The denominator of the right-hand-side of Eq.\ \ref{posterior_paraeters_general} is equal to:
\begin{equation}
\label{evidence_gen}
P(\mathcal{D}| \mathcal{N}) = \int \prod_k d w_k \,  P(\mathcal{D}|\{w_i \}, \mathcal{N}) P(\{w_i \}|\mathcal{N})
\end{equation}
and it  is called the evidence for the model $\mathcal{N}$.  On the other hand, from the Bayes' rule we have: 
\begin{equation}
\label{Posterior_para_general}
P(\mathcal{N}|\mathcal{D}) = \frac{P(\mathcal{D}| \mathcal{N}) P( \mathcal{N})}{P(\mathcal{D})}.
\end{equation}
For given data $\mathcal{D}$, the $P(\mathcal{D})$ is fixed. Moreover  if one assumes that there are no model preferences at the beginning of the analysis (the prior is uniform over neural networks of certain scheme and then uniform within particular model) i.e.\ $P( \mathcal{N}_1)=P( \mathcal{N}_2)=...$, then $P(\mathcal{N}|\mathcal{D}) \sim  P(\mathcal{D}| \mathcal{N})$ and that the evidence (\ref{evidence_gen}) can be used to rank the models. In practice 

In a typical situation the integrated function in the formula (\ref{evidence_gen}) is peaked at some configuration 
$\{w\}_{MP}$, so the evidence can be computed in the Hessian approximation \cite{MacKay_thesis}, 
\begin{equation}
P(\mathcal{D}|\mathcal{N}) \approx P(\mathcal{D}|\{w_i\}_{MP},\mathcal{N}) \underbrace{(2\pi)^\frac{p}{2} |A|^{-\frac{1}{2}}}_{Occam\,factor},
\end{equation}
where $p$ is the number of parameters, $A_{ij}= - \left. \nabla_{w_i}\nabla_{w_j} \ln P(\{w_i \}| \mathcal{D},\mathcal{N})\right|_{\{w_i \}=\{w_i \}_{MP}}$, and $|A|=\det A$ . In this case the evidence is proportional to the likelihood at the maximum  multiplied by the Occam factor, which penalizes too complex models. 

In a non-Bayesian analysis only the likelihood at the maximum is accessible and the comparison 
of non-nested models is not straightforward. Moreover when the number of parameters of the model increases then the maximum of the likelihood also grows -- the model with larger number of parameters can fit the data better. 
In the Bayesian approach the contribution from the Occam factor makes models which over-fit the data less likely. 
Therefore we expect that the best model should be characterized by a good predictive power.

\section{Artificial Neural Networks}
\begin{figure}[htb!]
 \centering
 \includegraphics[width=\textwidth]{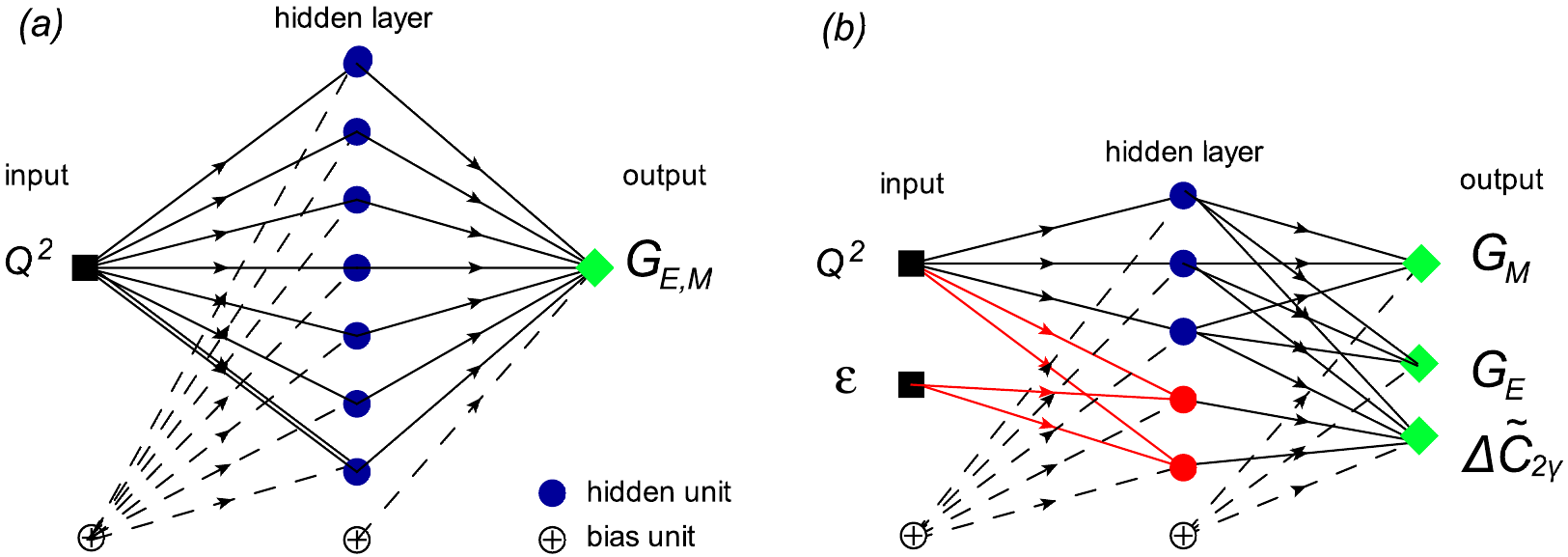}
 \caption{\label{Fig_Architecture} Fig. (a):
 MLP utilized to approximate the electric or magnetic FF of the nucleon \cite{Graczyk:2010gw}. It contains one input unit, one output unit, and one hidden layer with 7 (plus one bias) units. Fig. (b): MLP used to analyse  the elastic $e^\pm p$ data in order to extract the FFs of the proton and TPE correction \cite{Graczyk:2010gw}. MLP contains two input units, three output units and 5 (plus one bias) units in the hidden layer. The hidden layer is divided into FFs sector (blue vertices), which connects $Q^2$ input only with FFs, and TPE sector (red vertices), which does not connect with FFs. 
 Each line corresponds to one weight parameter. The  bias weights are denoted by dashed lines and the bias units by crossed circles.}
\end{figure}

In the particle and nuclear physics the artificial neural networks (ANNs) are used to  identify the interaction vertices and particles in the detectors \cite{NN_in_Physics}. Recently they are also exploited  to interpolate the parton distribution functions (PDFs) \cite{Ball:2013lla} and the nucleon form factors \cite{Graczyk:2010gw}. 

In the following two subsections we will review the foundations of the BF for feed forward neural networks \cite{MacKay_thesis,Bishop_book}.

\subsection{Multi-layer perceptron}

\begin{figure}[htb!]
 \centering
 \includegraphics[width=0.4\textwidth]{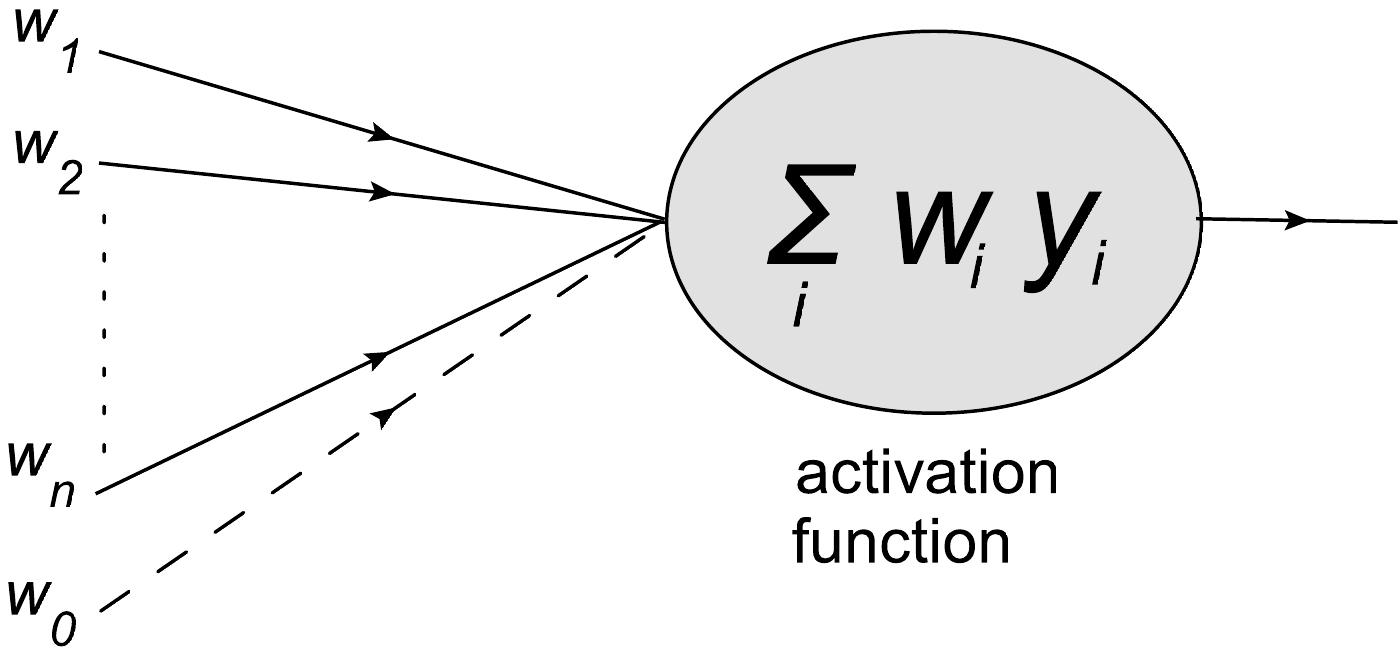}
 \caption{\label{Fig_Unit} Single unit connected with $n$ units (solid lines) and one bias (dashed line).
}
\end{figure}

In order to construct a statistical model for the function $\mathcal{N}$ we consider  feed-forward neural networks in multilayer perceptron (MLP) configurations\footnote{More detailed description of the MLP properties can be found in our previous paper \cite{Graczyk:2010gw} (Sect. 2).}. A neural network is a  non-linear map $\mathcal{N} : \mathbb{R}^{n_{in}} \to  \mathbb{R}^{n_{out}}$, where $n_{in/out}$ is the dimension of the input/output vector space,
which is usually represented as a graph with several layers of units (vertices) such that 
only the vertices from the consecutive layers can be connected. The first layer is  the input,
the last one is the output, and all other layers are hidden.
 Each unit (see Fig.\ \ref{Fig_Unit}) contains a real-valued function (called the activation function $f^{act}$) depending on one argument which is the weighted sum of the values obtained from the connected units from the previous layer,
\begin{equation}
y=f^{act}\left(\sum_{i\in \,{\rm previous\, layer}} w_{i} y_{i}\right).
\end{equation}
The weights $\{w_{i}\}$ (located at the edges of the graph) are real numbers, which are the parameters established during the training (learning) process, so as to maximize the posterior probability (\ref{posterior_w}).  A simple example of MLP, which was used to fit the electromagnetic proton, neutron FFs data is shown in Fig. \ref{Fig_Architecture} (a).

According to the Cybenko theorem \cite{Cybenko_Theorem}  the class of  networks with only one hidden layer containing sigmoid-like activation functions
and the output layer with linear activation functions  is dense in the space of continuous functions, $\mathbb{R}^{n_{in}} \to  \mathbb{R}^{n_{out}}$, defined on the unit hypercube.
This means that any continuous function can be approximated, with arbitrary precision,
by such a network, if it has sufficient number of units in the hidden layer. 
Therefore we restrict our analysis to networks of this class.

In the hidden layer we use sigmoid activation function $ f_{act}(x) = {1}/{(1 + \exp(-x))} $, which has effective support limited to a close neighbourhood of $x=0$. This property is used  when the prior for the weights is postulated and when the weights are randomly initialized at the beginning of each training process. 

\subsection{Bayesian framework for MLP}
To construct a statistical model the prior for the model parameters, weights, must be postulated. We consider a Gaussian distribution centred at zero.
\begin{eqnarray}
\label{prior_BF}
\mathcal{P}\left(\{w_i\}\right|\left. \alpha, \mathcal{N}\right)
&=& 
\frac{e^{- \alpha E_w}}{N} ,\; E_w=\frac{1}{2}\sum_{k=1}^{p}w_k^2, \; N  =  \int d^{p} w\, e^{- \alpha E_w}.
\end{eqnarray}

The parameter $\alpha$ (or rather the square root of its inverse), called later regularizer, defines the width of the Gaussian. If $\alpha$ is large then the prior (\ref{prior_BF}) dominates in the posterior (\ref{posterior_paraeters_general}), which forces the optimal weights $\{w_i\}_{MP}$ to be small. For low $\alpha$ the weights are unconstrained and the maximum of the likelihood dominates. As a result the optimal model may over-fit the data. The properly adjusted $\alpha$  prevents the over-fitting but does not affect strongly the obtained results, asserting that the model will have a good predictive power. Hence the prior  plays also the role of a penalty  contribution, which allows to prefer smaller rather than more complex  networks.

 In general one could introduce a separate regularizer for every weight. However, because of the internal symmetry of the MLP (hidden units from the same layer can be interchanged without affecting the value of the output) the set of weights can be divided in several distinct classes of parameters with one $\alpha$ parameter shared by all weights of a given class. For an instructive example see Sect. 3.2 of Ref.  \cite{Graczyk:2010gw}). However, to simplify our numerical calculations we consider only one common regularizer. 

 In principle $\alpha$ is one of the model parameters. Hence there should exist a configuration $\{\alpha, w_{i}\}_{MP}$, which maximizes expression (\ref{Posterior_para_general}). 
 In order to obtain their optimal values  together with the value of the evidence we use the so-called {\it evidence approximation} \cite{evidence}.

 The idea of this approach  is to  calculate  the posterior for weights,
 \begin{eqnarray}
\label{posterior_w}
P(\{w_i\}| \mathcal{D}, \alpha, \mathcal{N}) &=&
\frac{\mathcal{P}\left(\mathcal{D}\right|\left.\{w_i\}, \alpha, \mathcal{N} \right)
\mathcal{P}\left(\{w_i\}\right|\left. \alpha, \mathcal{N} \right)}{\mathcal{P}\left(\mathcal{D}\right|\left. \alpha, \mathcal{N} \right)},
\end{eqnarray}
assuming a fixed $\alpha$. It is done by taking the Gaussian (\ref{prior_BF}) for the prior and assuming that the likelihood is given by 
\begin{eqnarray}
\label{likelihood_BF}
\mathcal{P}\left(\mathcal{D}\right|\left.\{w_i\}, \alpha, \mathcal{N} \right)&=& 
\frac{e^{-\chi_{ex}(\mathcal{D},\{w_i\})}}{n},
\end{eqnarray}
where $\chi_{ex}(\mathcal{D},\{w_i\})$ is the $\chi$-square distribution for given data $\mathcal{D}$, and  $n$ -- the normalization factor calculated in the Hessian approximation, see Eq. (3.8) of Ref. \cite{Graczyk:2010gw}.

In this approximation the maximum of the posterior (\ref{posterior_w}) (for fixed $\alpha$) corresponds to the minimum of  the error function:
\begin{equation}
\label{minimum_of_error}
S(\mathcal{D},\alpha,\{w_i\}) = \chi^2_{ex}(\mathcal{D},\{w_i\}) + \alpha E_w.
\end{equation}
Hence the optimal configuration of weights minimizes $S(\mathcal{D},\alpha_{MP},\vec{w})$.

On the other hand, it can be shown \cite{Bishop_book} that the necessary condition for the optimal $\alpha$ reads
\begin{equation}
\label{alpha_iter}
\left.\frac{\partial }{\partial \alpha} \mathcal{P}\left(\mathcal{D}\right|\left. \alpha, \mathcal{N} \right)\right|_{\alpha=\alpha_{MP}} =0.
\end{equation}
(notice that $ \mathcal{P}\left(\mathcal{D}\right|\left. \alpha, \mathcal{N} \right) = \int d^p w \mathcal{P}\left(\mathcal{D}\right|\left.\{w_i\}, \alpha, \mathcal{N} \right)
\mathcal{P}\left(\{w_i\}\right|\left. \alpha, \mathcal{N}\right) $). 

Both conditions (\ref{minimum_of_error}) and (\ref{alpha_iter}) are used to  find (iteratively) the optimal configuration $\{\alpha, w_{i}\}_{MP}$. Then  the evidence for the model $P(\mathcal{D}|\mathcal{N})$ is computed. It is done by using the results of the previous step, for the details see \cite{Bishop_book} (Chap. 10).

\section{Application: Investigation of the Proton Structure}

 The feed-forward neural networks can be used to approximate the FFs \cite{Graczyk:2010gw}. The methods of neural networks allow one to reduce the model-dependence of the results of the analysis  and to make predictions in kinematic regions where there are no measurements. 
 
 As an example of application to hadron physics we present the extraction of the E-M proton FFs and the TPE correction from the elastic $e^\pm p$  scattering data. Our aim  is to discuss the statistical features of the approach like dealing with the bias-variance trade-off  and the problem of estimating of the systematic uncertainty due the 
 model-dependence. 

\subsection{Form-factors and two-photon exchange correction}

A part of the information about the structure of the nucleon  is hidden in  the E-M FFs \cite{Belitsky:2003nz}. They are the  functions which parametrize the E-M proton vertex \cite{Thomas_book}:
\begin{equation}
\Diagram{ &  \vertexlabel^{q} \\
&  ![ulft]{gv}{} &  \\
{fA} & ![bot]{fA}{} \\
} = \Gamma^\mu(q) = F_1(Q^2) \gamma^\mu + \frac{\mathrm{i}\sigma^{\mu\nu}q^\nu }{2M_p} F_2(Q^2),
\end{equation} 
where $M_p$ is the proton mass, $q$ is the four-momentum of the virtual photon, and $F_{1,2}$ denote the form-factors -- Lorentz invariant scalars depending only  on the four-momentum transfer $Q^2=-q^2$.   

It is convenient to consider the electric, $G_E = F_1 - (Q^2/4M^2_p)F_2$, and the magnetic, $G_M = F_1+F_2$, FFs which  at low $Q^2$ and in the Breit frame can be related with the Fourier transform of the charge and the magnetic distributions inside the proton \cite{Ernst:1960zza}. 

The E-M FFs are extracted from the elastic $e^- p$ scattering data within two methods \cite{form_factors_review}. In one method the Rosenbluth separation of the  unpolarized cross section data is performed, and the values of $G_E$ and $G_M$ are simultaneously obtained.  In the other method (denoted later as PT)  the ratio $G_E/G_M$ is extracted from the measurements of the polarization transfer observables. In both types of the analyses the radiative corrections  are subtracted from the scattering data in order to get the cross sections, polarizabilities, in \textit{the one-photon exchange approximation}. 

It turned out that the $G_E/G_M$ ratios obtained from the Rosenbluth separation and from the PT measurements differ at larger $Q^2$ values (for the review see \cite{Arrington:2011dn}). This inconsistency can be partially removed if the Rosenbluth data are corrected by the TPE contribution (hard photon part)\cite{tpe_wyjasnienie,Arrington:2007ux}, which was neglected in the old analyses.  It is a small correction but its inclusion into the Rosenbluth analysis changes the results of the separation (affecting mainly the resulting value of $G_E$), while the form factor ratio obtained from the PT data analysis is affected to much lesser extent.

\begin{figure}[htb!]1
 \centering
\includegraphics[scale=0.6]{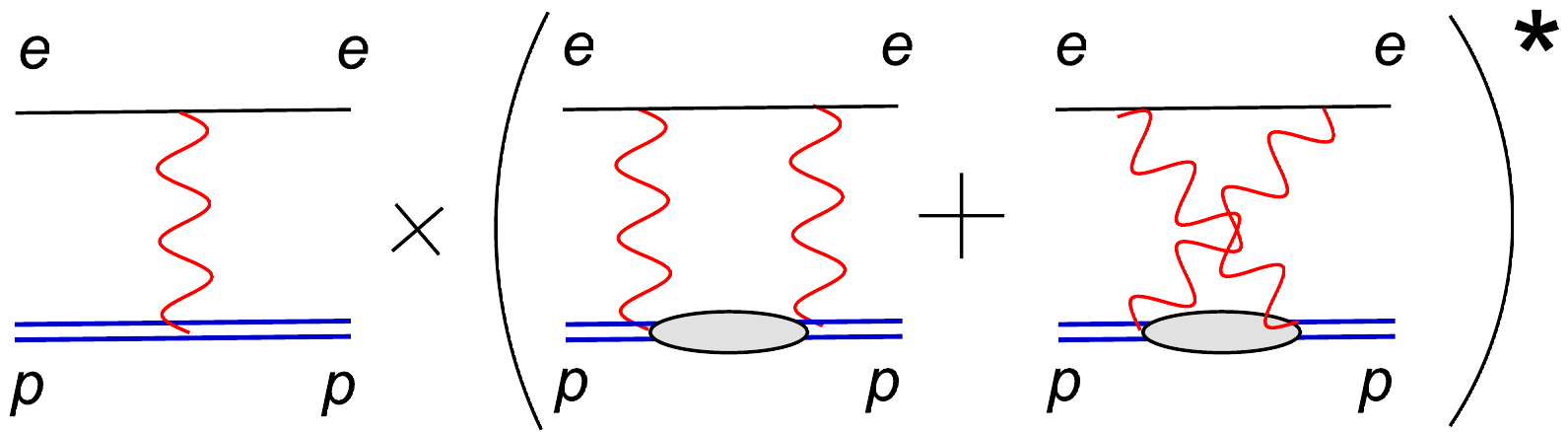}
\caption{Two-photon exchange contribution to elastic $ep$ scattering. \label{Fig_tpe}}
\end{figure}

The TPE  correction is  given by the interference of the Born diagram and  the diagrams describing the exchange of two virtual photons between the electron and the proton (Fig. \ref{Fig_tpe}).
 The hard photon contribution of the TPE is induced by the hadronic  proton structure. Many efforts have been made to calculate this contribution based on the  theoretical and phenomenological  models (see references 6-18 in \cite{Graczyk:2011kh}). However, for larger $Q^2$ values the predictions of TPE are model-dependent. On the other hand, there are attempts to  get the TPE contribution directly from the scattering data, by using constructively the inconsistency of Rosenbluth-PT data. Obviously such analysis requires  model assumptions about the functional form of the FFs and the TPE term. However, the extracted value of the TPE correction depends also on the choice of the model i.e.\ the FFs and the TPE parametrizations. The BF for MLP turned out to be a useful  methodology for dealing with the model-dependence and finding the optimal statistical model \cite{Graczyk:2011kh}. In the next subsections we review the main features of this approach. 

The TPE correction is particularly well suited to discuss model-dependence because
both the extraction of the TPE correction from the data  and also its theoretical predictions are affected by the choice of the model. On the other hand recently the ratio $R_{+/-} = \sigma(e^+p)/\sigma(e^-p)$ of the cross sections for elastic positron-proton to the electron-proton scattering is  measured in two dedicated experiments \cite{Gramolin:2011tr,Bennett:2012zza}. From (\ref{ratio_R+-}) it is evident that the TPE correction contributes to the deviations of  $R_{+/-}$ from unity.

\subsection{Analysis of the elastic $e^\pm p$ scattering data}

For the purpose of this presentation we made a revision of our previous analysis \cite{Graczyk:2011kh}. The present results are obtained by an improved version of the neural network program   with more efficient learning algorithm  (Levenberg-Marquardt \cite{LMalogorithm}). The structure of the program has also been changed. It is now possible to consider larger number of models in shorter time.  

The idea of the analysis is to assume that the missing correction, responsible for the Rosenbluth-PT data disagreement, affects mainly the cross section data and the PT observables to much lesser extent \cite{tpe_wyjasnienie}. Hence performing a combined analysis of both types of the data should allow one to get  ''missing'' contribution responsible for the disagreement.  In practice we consider three types of the data:
\begin{enumerate}
\item unpolarized cross sections (27 independent data sets);
\item the $G_E/G_M$ ratio (from PT measurements); 
\item the $R_{+/-}$ ratio.
\end{enumerate}
All three types of data depend on $(G_E,G_M, \Delta \tilde{C}_{2\gamma})$, where $\Delta \tilde{C}_{2\gamma}$ denotes the ''missing'' correction responsible for the disagreement. $\Delta \tilde{C}_{2\gamma}$ contributes to the reduced unpolarized cross section: 
\begin{equation}
\label{sigma_reduced}
\sigma_{R}(Q^2, \varepsilon,s) = \frac{Q^2}{4M^2_p} G_M^2(Q^2) + \varepsilon G_E^2(Q^2) +   s\Delta \tilde{C}_{2\gamma}(Q^2, \varepsilon),
\end{equation}
and it is interpreted as TPE contribution. In (\ref{sigma_reduced}) $s=\pm 1$ corresponds to the cross-section for $e^\pm p $ scattering and $\varepsilon = \left[1+2\left(1+{Q^2}/{4M_p^2}\right)\tan^2\!\left({\theta}/{2}\right)\right]^{-1}$ is the photon polarizability ($\theta$ is the scattering angle between incoming and outgoing electrons).
It is easy to see that the ratio of the positron/electron cross-sections has the form:
\begin{equation}
\label{ratio_R+-}
R_{+/-} = \sigma_{R}(Q^2, \varepsilon,-1)/\sigma_{R}(Q^2, \varepsilon,+1).
\end{equation}

In the combined analysis of the data we consider  networks with two inputs $(Q^2,\varepsilon)$ and three outputs 
$(G_E,G_M, \Delta \tilde{C}_{2\gamma})$, see Fig. \ref{Fig_Architecture}. Because the FFs do not depend on $\varepsilon$ some of the connections in the network are erased. As a result the hidden layer of the network is divided into two parts: one (called FF sector) contains $g$ units connected only with  $Q^2$ input and both outputs, and the other (called TPE sector) with units disconnected only from the TPE output. 

The likelihood (\ref{likelihood_BF}) and prior (\ref{prior_BF}) are defined in the same way as in our previous analysis: Eqs. 10, 11 A1-3 and 12, 13  of Ref. \cite{Graczyk:2011kh} respectively. The  selection of the data sets is the same as well. The $\chi^2_{ex}$ in (\ref{likelihood_BF}) is the sum of the $\chi^2$s' for the cross section, FFs ratio and $R_{\pm}$ data. The experimental data points are characterized by statistical and point to point systematic uncertainties.
% added in the quadrature. 
In the case of the cross section data, for every separate set of measurements, the systematic normalization uncertainty is taken into account and corresponding   normalization parameter is introduced into the fit, for more details see Appendixes A and B of Ref. \cite{Graczyk:2011kh}.

\subsection{Numerical Algorithm}

The scheme of the numerical analysis is the following:
\begin{enumerate}
\item consider MLP with definite number of hidden units;
\item find the optimal configuration of weights $\{w_i\}_{MP}$ and $\alpha_{MP}$ by the use of a learning  algorithm:
\begin{enumerate}
\item randomly initialize the weights;
\item perform the learning trial, iterating on-line $\alpha$ parameter (as described above);
\end{enumerate}
\item compute the evidence for the model (for the analytic expression see Eqs. 33-35 of \cite{Graczyk:2013pca});
\item for a given network type (scheme) choose the best model (one with  the highest evidence);
\item change the network type (by increasing by one the number of units either in form-factor or TPE part of the hidden layer) and repeat the steps (i-v).
\end{enumerate}

\begin{figure*}[h]
\centering{
\includegraphics[width=0.80\textwidth]{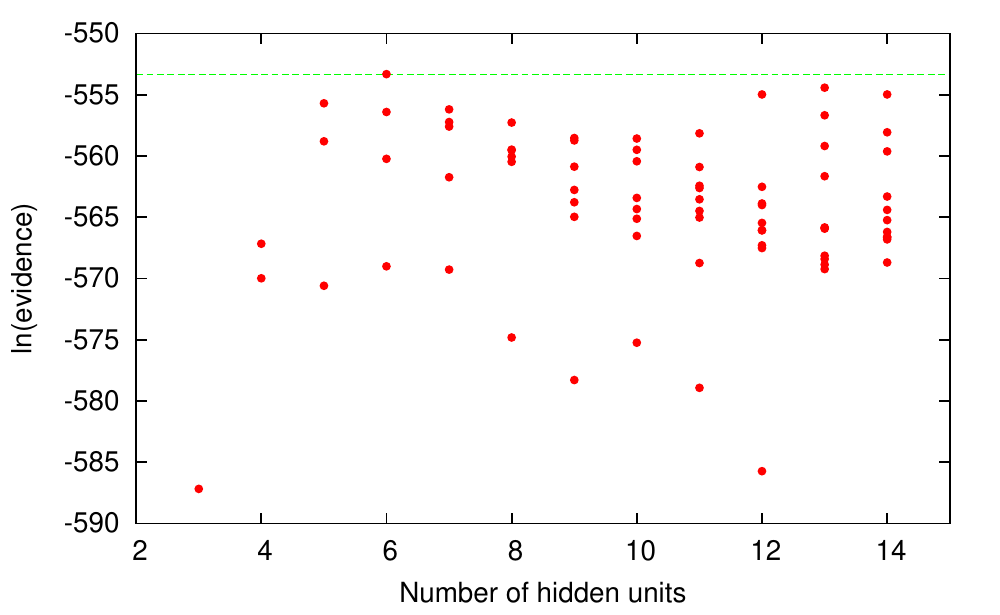}
}
\caption{Dependence of the logarithm  of evidence on the total number of hidden units. 
For each total number of hidden units there are many models with different numbers of units in the FF and TPE sectors.
\label{Fig_evidence}}
\end{figure*}

We considered 102 different MLP schemes. The maximal number of units in the hidden layer was 14. For each type of the network about  2300 learning processes were performed reaching  the total number  of about 226 000 considered networks. After that we collected distribution of 102 models ranked by the evidence $P(\mathcal{D}|\mathcal{N})$. The model with the highest evidence contains 6 hidden units: 2 in the FF sector and 4 in the TPE sector (see Fig. \ref{Fig_evidence}).

\subsection{Searching for the optimal model}

\begin{figure*}[h]
\centering{
\includegraphics[width=\textwidth]{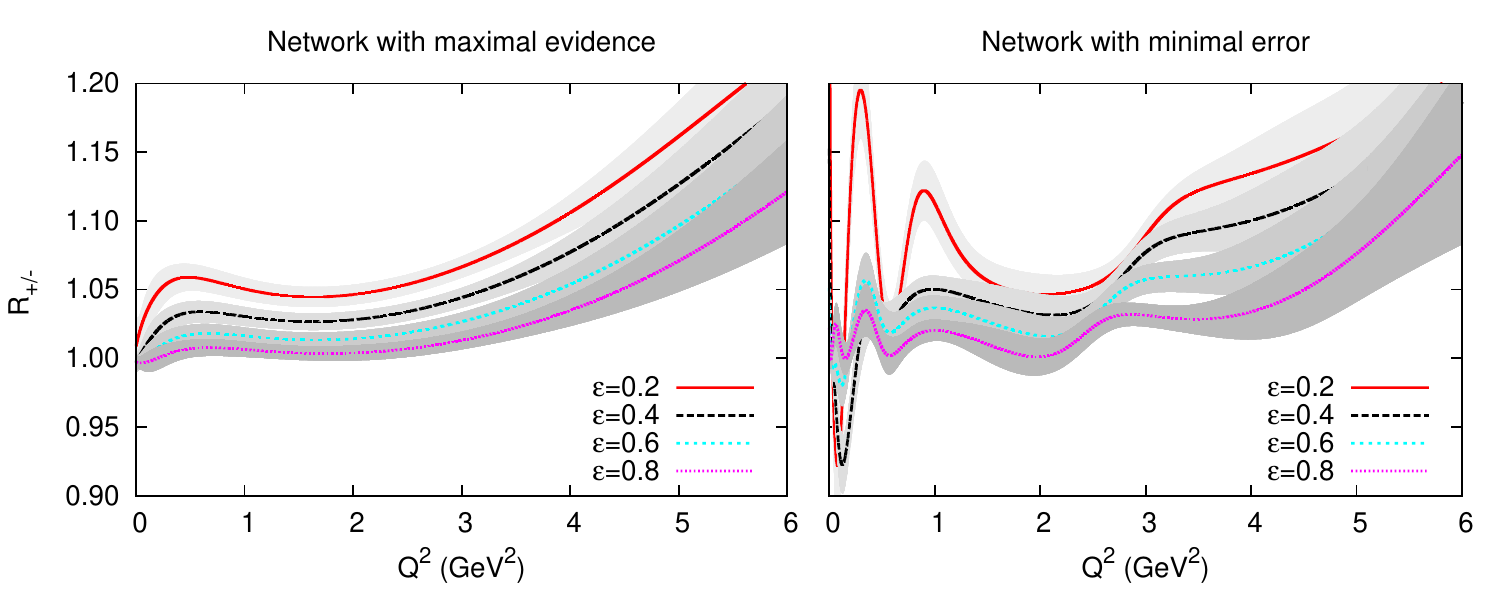}
}
\caption{\label{Fig_R_+-_best_evidence_error}   Left plots:
predictions of the ratio $R_{+/-}$ based on  the model which 
maximizes the evidence, $\ln(P(\mathcal{D}|\mathcal{N}))=-547$. Right plots: predictions of the ratio $R_{+/-}$ based on model, which minimizes  the error function, here  $\ln(P(\mathcal{D}|\mathcal{N}))=-646$.}
\end{figure*}

The optimal model, which we search for, should be rather simple (low number of parameters) to have ability for generalization (making the predictions about new data). On the other hand the number of parameters should be large enough so that model  be able to reproduce the current data with reasonable precision.  These two requirements are opposite, which is called the bias-variance trade-off. The optimal solution  is a compromise between both tendencies  \cite{Bishop_book}. 
\begin{figure*}[h]
\centering{
\includegraphics[width=0.8\textwidth]{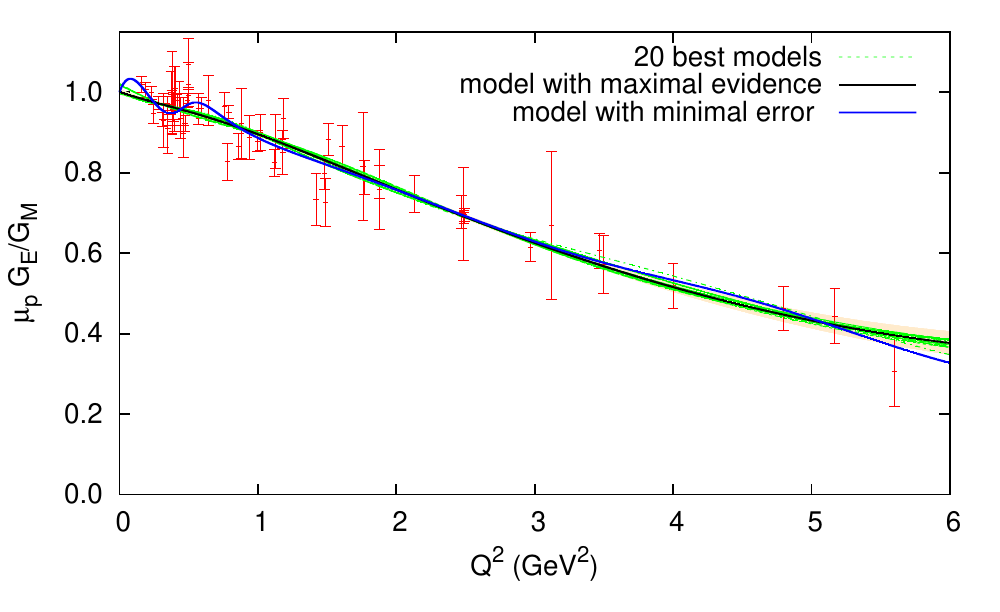}
}
\caption{Polarization transfer $\mu_p G_E/G_M$ ratio data  against the best fit, 20 best models and the model which overfits the data ($\mu_p$ is the proton magnetic moment in the units of the nuclear magneton). 
\label{Fig_FF_ratio}}
\end{figure*}

In the BF finding the optimal solution is achieved in a very natural way. As  mentioned before, the Bayesian statistics embodies Occam's razor. In our approach this appears in two ways: first, by adjusting value of the regularizer $\alpha$, second, by using the evidence to choose the best model.

The most spectacular differences between the predictions of the model  which is  optimal (maximizes the evidence)  and the one, which minimizes the error function and over-fits the data  are seen in the plot of the $R_{+/-}$ (Fig. \ref{Fig_R_+-_best_evidence_error}). Similar comparison for the $\mu_p G_E/G_M$ ratio is shown in Fig.\ \ref{Fig_FF_ratio}. 
The overfitted  parametrization is characterized by high curvatures, while the optimal parametrization is given by much smoother curves.  
In this case the example of over-fitted parametrization is rather spectacular, it minimizes the error function, and from the point of view of $\chi^2/NDF$ is highly acceptable. The only reason (in qualitative sense) for rejecting this model from the typical non-Bayesian analysis is the presence of unacceptably high curvatures. But in a less spectacular case such model would be accepted. In the BF the best model is indicated by a mathematical objective algorithm i.e.\ the way of learning the networks and the evidence hence the ''human'' decision is reduced to a minimum.

\subsection{Systematic uncertainty induced by choice of the parametrization}
\begin{figure*}[h]
\centering{
\includegraphics[width=0.9\textwidth]{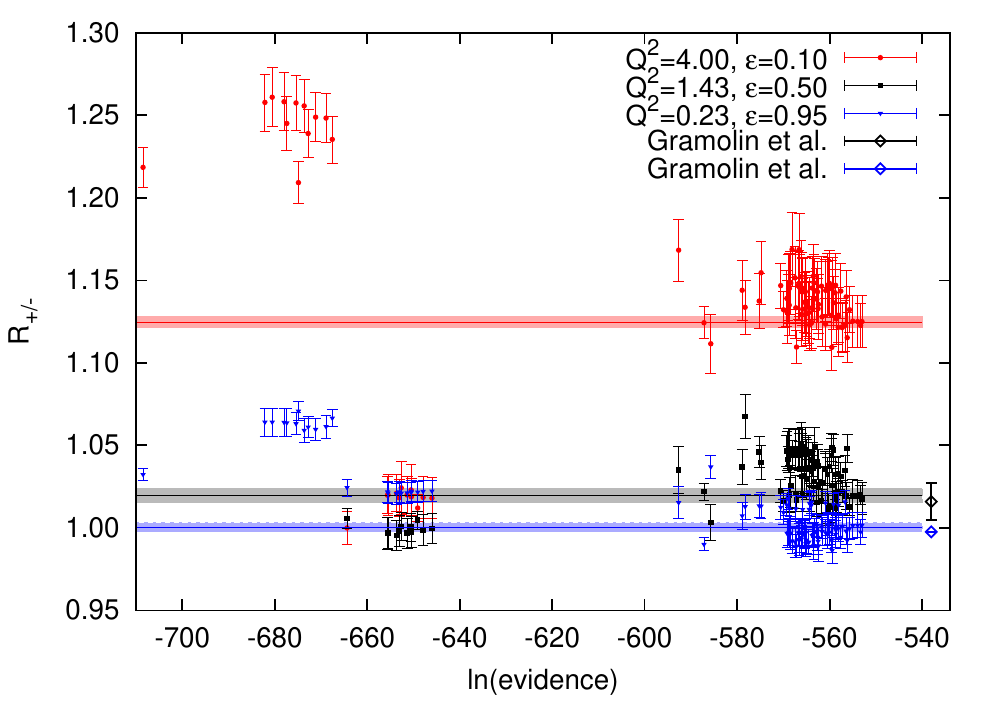}
}
\caption{The  ratio $R_{+/-}$ calculated for three points $(Q^2,\varepsilon)=$ $(4 \,\mathrm{GeV}^2,0.10)$, $(1.43\, \mathrm{GeV}^2,0.50)$ and $(0.23\, \mathrm{GeV}^2,0.95)$, in the last two points correspond to the recent  measurements done by Gramolin \textit{et al.} \cite{Gramolin:2011tr}, denoted here by open diamonds. Each single point denotes the value of $R_{+/-}$ calculated for one particular model, in this case the error bounds are related with the weights uncertainty. The solid lines show the mean values calculated from (\ref{discrete_series}). The shaded area denotes $1\sigma$ uncertainty due to the model-dependence of the parametrizations. 
\label{Fig_systematic_3_poins}}
\end{figure*}

The main result of the analysis  is the set (denoted by $\mathcal{M}$)  of the neural networks, parametrizations, which are ranked by the evidence. Each of them has a different  connection graph and maximizes the evidence in its class of functions. Having such distribution of models one can estimate the systematic uncertainty due to the choice of the functional parametrization (the network shape).

Let us introduce the mean value, according to the space of \textit{the best parametrizations} of the observable $\mathcal{F}$, which is a function of outputs of the network $\mathcal{N}$,
\begin{eqnarray}
\label{discrete_series}
\overline{\mathcal{F}(G_E,G_M, \Delta
\tilde{C}_{2\gamma}) } &=& 
\sum_{\mathcal{N}\in \mathcal{M}}
\mathcal{F}(G_E^{\mathcal{N}},G_M^{\mathcal{N}}, \Delta \tilde{C}_{2\gamma}^{\mathcal{N}}) \mathcal{P}_{nor}(\mathcal{D}|\mathcal{N}),
\end{eqnarray}
where, 
\begin{equation}
\mathcal{P}_{nor}(\mathcal{D}|\mathcal{N}) = \frac{\mathcal{P}(\mathcal{D}|\mathcal{N})}{\sum_{\mathcal{N}\in \mathcal{M}} \mathcal{P}(\mathcal{D}|\mathcal{N})}
\end{equation}
and the relation $\mathcal{P}(\mathcal{D}|\mathcal{N}) \approx \mathcal{P}(\mathcal{N}|\mathcal{D})$ is also imposed.

Then the systematic uncertainty due to the choice of the functional parametrization is given by the square root of the variance $\Delta \mathcal{F}(G_E,G_M, \Delta
\tilde{C}_{2\gamma})$. 

In Fig.\ \ref{Fig_systematic_3_poins} we plot the estimates of the 
$R_{+/-}$ (for three points) against the logarithm of evidence. Each point denotes the value predicted by a neural network with the highest evidence in its class. The prediction is plotted  together  with the error bounds ($1\sigma$ error due the distribution of weights) see Eqs. 3.15 and 3.16 of \cite{Graczyk:2010gw}. The horizontal lines denote the mean value (weighted by the evidence) see (\ref{discrete_series}), while the shaded areas denote the $1\sigma$ uncertainty given by  $\Delta R_{+/-}$, calculated as described above. Notice that the best model predictions agree well with the new data from \cite{Gramolin:2011tr}, which were not included in the analysis. On the other hand it is clear that many models give negligible contribution to the mean value and dispersion because they have very low values of the evidence. 

%It turned out that the systematic uncertainty is very small for $\mu_p G_E/G_M$ ratio, see Fig. \ref{Fig_FF_ratio}, where we plot the best fit together with 20 best models according to the evidence.  
\begin{figure*}[h]
\centering{
\includegraphics[width=0.8\textwidth]{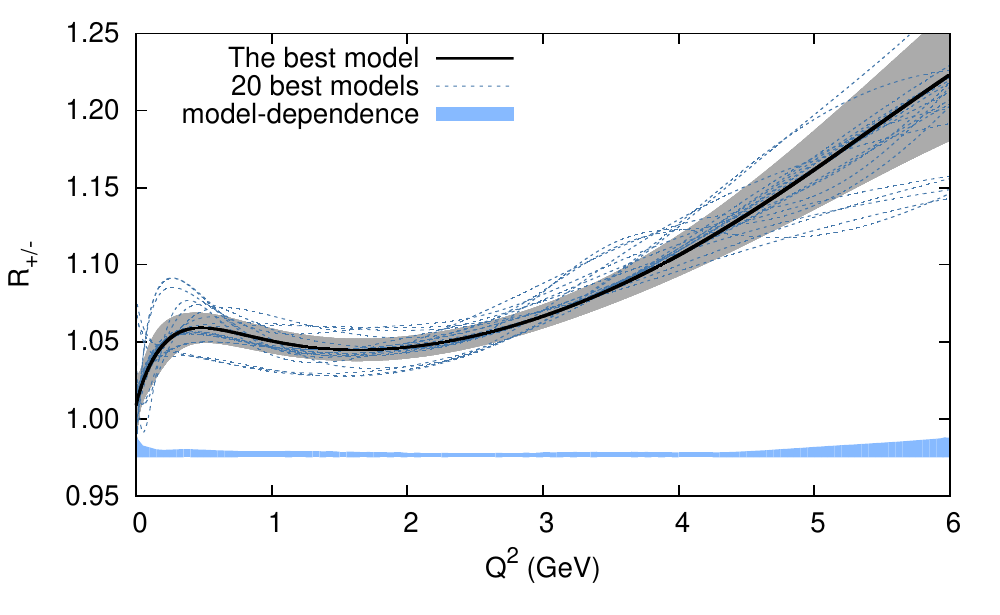}
}
\caption{The  ratio $R_{+/-}$, the best model and best 20 fits according to evidence. The grey area denotes the $1\sigma$ uncertainty due to the distribution of the weights, while the blue area denotes the systematic uncertainty, due to the choice of the functional parametrization. 
\label{Fig_systematic_plot}}
\end{figure*}
In Fig. \ref{Fig_systematic_plot} we present another example of estimating the systematic uncertainty due to the model-dependence. It is the $Q^2$ dependence of ratio $R_{+/-}$ calculated for a  fixed value of $\varepsilon$ . The systematic uncertainty is very small on a large $Q^2$ range. In the same figure for the qualitative comparison we plot also the best 20 fits, due to the evidence.

\begin{figure*}[h]
\centering{
\includegraphics[width=0.8\textwidth]{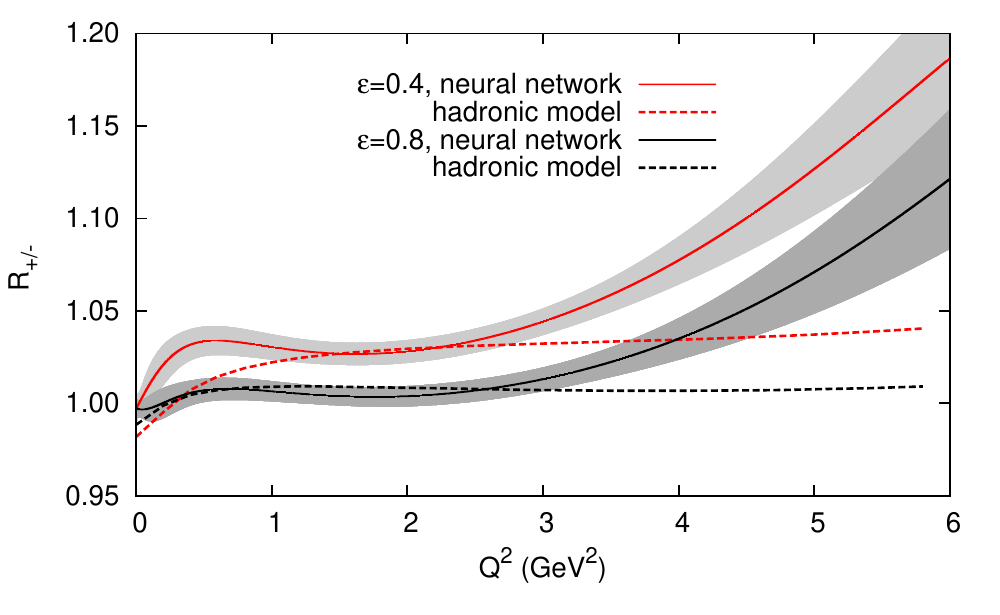}
}
\caption{The  ratio $R_{+/-}$ for different $\varepsilon$ values estimated based on the best neural network model and the hadron model from \cite{Graczyk:2013pca}. The shaded areas denote the $1\sigma$ uncertainty due to the distribution of the weights.
\label{Fig_NN_th}}
\end{figure*}

Obviously the above estimate of the  systematic model-dependence uncertainty does not include all the model dependence of the approach. Indeed, an important assumption to perform the analysis was to neglect the TPE correction to the PT data. This model-assumption is difficult to quantitatively account for. One can only estimate the systematic bias with respect to the theoretical model calculations.  In Fig. \ref{Fig_NN_th} we compare the predictions of $R_{+/-}$ obtained by neural network  with hadronic model predictions \cite{Graczyk:2013pca}. In the latter  the TPE is calculated within the quantum field theory approach,  in which it was assumed that the hadronic intermediate state is given by either a proton or the $P_{33}(1232)$ resonance \cite{Graczyk:2011kh}. It is expected that this model should work well in the low and intermediate $Q^2$ range. It can be seen that in the low $Q^2$ there is a systematic deviation between the neural network response and the theory. It seems that it is the result of the theoretical assumption mentioned above. On the other hand it can also be caused by the inaccuracy of the theoretical model.

\section{Summary}

We have shown that the approach based on neural networks can be used to extract important  information about the structure of the proton from the scattering data. The approach  offers tools which make it possible to control and reduce the model-dependence. The result of the analysis is a statistical model with good predictive power which can be used to make predictions about the FFs and TPE correction in the kinematic region where there are no measurements.

It seems that one can try to introduce an analogical Bayesian framework also in the case of non-neural network analyses, especially,  for the set of theoretical models. However, it seems that in that case the construction of an objective prior distribution is a challenge. Additional difficulty is the over-fitting problem and construction of a suitable penalty term.

\section*{Acknowledgements}

The calculations have been carried out at the Wroclaw Centre for Networking and Supercomputing (\url{http://www.wcss.wroc.pl}), grant No. 268.

\end{document}